%% file: icrc2023_MeV.tex
\documentclass[a4paper,11pt]{article}
\usepackage{pos}
\usepackage{wrapfig}
\usepackage{physics}
\usepackage{lineno}

\begin{document}

\title{Constraining MeV Neutrino Emission of Bright Transients with IceCube
}
 \ShortTitle{MeV Neutrinos with IceCube}

\author{The IceCube Collaboration \\{\normalsize \normalfont(a complete list of authors can be found at the end of the proceedings)}\\}





\emailAdd{nora.valtonen-mattila@icecube.wisc.edu}
\emailAdd{sgriswold@icecube.wisc.edu}
\emailAdd{sybenzvi@icecube.wisc.edu}

\abstract{MeV neutrinos are produced in many astrophysical transients, such as stellar collapses and high-energy jets, where they play a role in sustaining and cooling energetic explosions. Detecting these neutrinos from sources outside the Milky Way is very difficult due to the small neutrino-nucleon cross section at MeV. Nevertheless, the non-observation of MeV neutrinos from high-energy transients may provide useful constraints on related neutrino production mechanisms where significant MeV production is expected. The IceCube Neutrino Observatory, a cubic kilometer neutrino detector operating with nearly 100\% uptime at the South Pole, is sensitive to bursts of MeV neutrinos from astrophysical sources in and beyond the Milky Way. In this work, we describe the MeV neutrino detection system of IceCube and show results from several categories of astrophysical transients.

\vspace{4mm}
{\bfseries Corresponding authors:}
Segev BenZvi$^{1*}$, Spencer Griswold$^{1}$, Nora Valtonen-Mattila$^{2}$\\
{$^{1}$ \itshape 
Department of Physics and Astronomy, University of Rochester, Rochester, NY 14627, USA
}\\
{$^{2}$ \itshape 
Department of Physics and Astronomy, Uppsala University Box 516, S-75120 Uppsala, Sweden
}\\
$^*$ Presenter

\ConferenceLogo{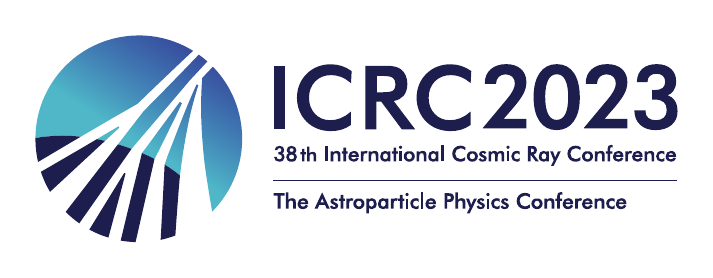}

\FullConference{%
The 38th International Cosmic Ray Conference (ICRC2023)\\
  26 July -- 3 August, 2023\\
  Nagoya, Japan}
}

\maketitle

\section{Detecting MeV Neutrinos at the IceCube Neutrino Observatory}\label{sec:mev_detection}

The IceCube Observatory, a 1~km$^3$ hexagonal lattice of photomultiplier tubes (PMTs) buried in the deep Antarctic ice beneath the geographic South Pole, detects the Cherenkov photons produced when neutrinos interact in the ice and create secondary charged particles. The detector consists of 86 vertical cables, or strings, spaced an average of 125 m apart~\cite{IceCube:2016zyt}. Each string is instrumented with 60 digital optical modules (DOMs) which encapsulate the PMTs and their associated readout electronics 
in spherical glass pressure vessels. The DOMs are sensitive to wavelengths between 300~nm and 600~nm and are deployed at depths of 1450~m to 2450~m in the ice with a vertical separation of 17~m. A centrally located group of eight out of 86 strings forms the DeepCore subarray, which has an average inter-string spacing of 72~m and an inter-DOM spacing ranging from 7~m to 10~m \cite{IceCube:2011ucd}. 

\begin{wrapfigure}{r}{0.5\textwidth}
    \includegraphics[width=\linewidth]{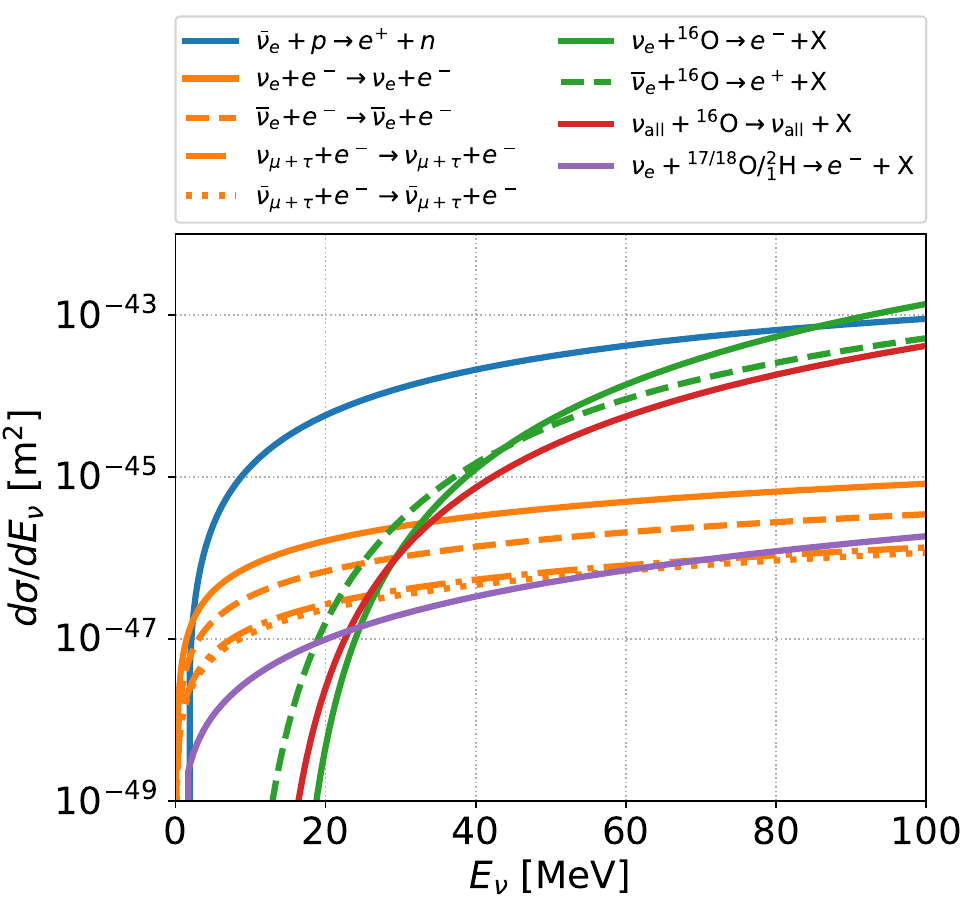}
    \caption{Neutrino interaction cross sections in ice below 100 MeV.}
    \label{fig:mev_neutrino_xs}
\end{wrapfigure}

The spacing of the DOMs is optimized to observe neutrinos with energies ranging from approximately 5~GeV to 1~PeV and beyond, though advanced analysis techniques can lower the energy threshold to sub-GeV \cite{IceCube:2023elowen}. In addition, IceCube is sensitive to bursts of MeV neutrinos lasting a few seconds. Below 50 MeV, the dominant neutrino interaction in the ice is inverse beta decay (IBD: $\bar{\nu}_e+p\to e^+ + n$), with percent-level contributions from all-flavor neutrino-electron elastic scattering (Fig.~\ref{fig:mev_neutrino_xs}). Above 50~MeV, charged-current interactions of $\nu_e$ and $\bar{\nu}_e$ with $^{16}$O nuclei, and neutral-current interactions of all flavors with oxygen, provide increasingly important contributions to neutrino interactions in the ice \cite{IceCube:2011cwc}.

To understand the sensitivity of IceCube to MeV neutrinos, it is instructive to explore the inverse beta process which dominates interactions below 50~MeV. The positron produced in IBD travels an average distance of $0.56~\mathrm{cm}\times(E_{e^+}/\mathrm{MeV})$ in the ice and yields $\lesssim180$ Cherenkov photons per MeV between 300 and 600~nm \cite{IceCube:2011cwc}. The subsequent capture of the IBD neutron produces a 2.2~MeV gamma ray which Compton scatters electrons and creates $\sim100$ additional photons \cite{IceCube:2011cwc}. While the Cherenkov yield from single IBD interactions is $\mathcal{O}(10^3)$ photons, the significant difference between the path lengths of the  $e^+$ and $n$ final-state particles and the inter-DOM spacing of IceCube and DeepCore means that most IBD photons are scattered and absorbed in the ice before reaching a DOM. On average, a given IBD interaction will produce one recorded photoelectron in one DOM \cite{IceCube:2011cwc}. Thus it is not possible to reconstruct the arrival direction or energy of individual $\bar{\nu}_e$ events at MeV. Moreover, the photoelectrons from neutrino interactions must be detected on top of a per-DOM background rate of 300~Hz caused by triboluminescence in the ice and intrinsic radioactivity in the glass pressure vessels. Additional backgrounds of 3~Hz to 30~Hz per DOM, varying by season and by depth, are caused by Cherenkov light from atmospheric muons.

\section{Online Detection of MeV Neutrino Bursts}\label{sec:online_system}

\begin{figure}[ht]
    \centering
    \includegraphics[width=0.9\textwidth]{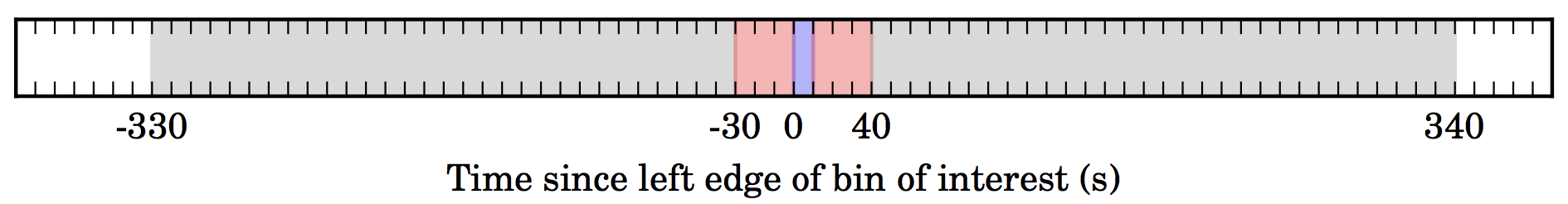}
    \caption{Sliding window excess count search. Counts in the signal window (blue) and background windows of $\pm300$~s (gray) are used to estimate $\Delta\mu$ in eq.~\eqref{eq:delta_mu}. Counts $\pm30$~s around the signal window (pink) are excluded.}
    \label{fig:sndaq_time_binning}
\end{figure}

The ever-present radioactive and muon backgrounds in the IceCube DOMs make it impractical to identify steady or slowly varying sources of MeV neutrinos. However, the large size of the detector means it is sensitive to correlated bursts of many MeV neutrinos arriving on timescales of $\mathcal{O}(1~\mathrm{s})$. In lieu of reconstructing $\bar{\nu}_e$ events, a dedicated pulse counter-based data acquisition system (DAQ) is used to search for bursts of MeV neutrinos by analyzing the count rates of all DOMs in real-time. An online algorithm records the DOM count rates in 2~ms time bins and then resamples the rates of individual DOMs, $r_i$, to bins of width 0.5~s. Taking each time bin as a search window for a neutrino burst, the algorithm computes the average count rate per DOM, $\expval{r_i}$, and the uncertainty in the average, $\sigma_i$, using sliding time windows of length 300~s before and after the search bin (see Fig.~\ref{fig:sndaq_time_binning} and \cite{IceCube:2011cwc}). The search algorithm then computes a test statistic $\xi = \Delta\mu / \sigma_{\Delta\mu}$ to identify MeV neutrino bursts, where
\begin{align}\label{eq:delta_mu}
    \Delta\mu &= \sigma_{\Delta\mu}^2\sum_{i=1}^{N_\mathrm{DOM}}\frac{\epsilon_i\qty(r_i-\expval{r_i})}{\sigma_i^2}
    &
    &\mathrm{and}
    &
    \sigma_{\Delta\mu} &= \qty(\sum_{i=1}^{N_\mathrm{DOM}}\frac{\epsilon_i^2}{\sigma_i^2})^{-1}.
\end{align}
Here $\Delta\mu$ is the maximum-likelihood estimator of the collective rate increase across all DOMs, weighted by the relative detection efficiency $\epsilon_i$ of each DOM \cite{IceCube:2011cwc, IceCube:2023sn}.

\begin{wrapfigure}{l}{0.48\textwidth}
    \includegraphics[width=\linewidth]{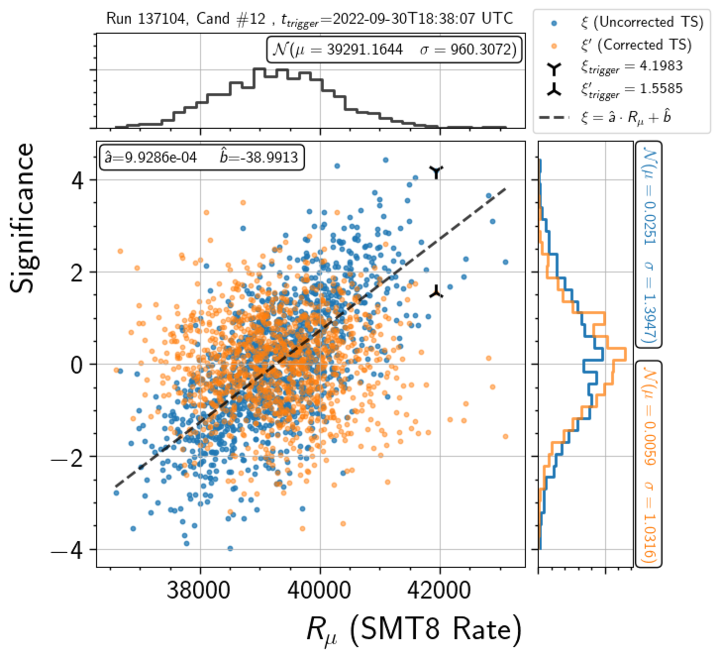}
    \caption{Count excess test statistic $\xi$ and muon-corrected test statistic $\xi'=\xi_\text{corr}$.}
    \label{fig:rate_corr}
\end{wrapfigure}

The distribution of the test statistic $\xi$ has large tails due to the seasonal effect of atmospheric muons, but this effect can be removed by tracking the correlation of $\xi$ with the real-time muon rate recorded by the IceCube simple multiplicity trigger \cite{Baum:2015drl}. The muon-corrected test statistic, $\xi_\mathrm{corr}$, is normally distributed (see Fig.~\ref{fig:rate_corr}). Large values of the corrected test statistic ($\xi_\mathrm{corr}\geq 5$) occur approximately once per month.

When $\xi_\mathrm{corr}\geq5$, corresponding to a notable collective increase in the count rate $\Delta\mu$, the online DAQ sends an automatic alert to experts within the collaboration as well as external monitoring networks such as the SuperNova Early Warning System \cite{SNEWS:2020tbu, IceCube:2023sn}. In addition, a large value of $\xi_\mathrm{corr}$ will trigger an automatic readout of the buffered DOM waveforms from the computers in the IceCube Laboratory on the surface of the ice sheet \cite{Baum:2013ekr}. The buffered readout saves untriggered waveforms from $-30$~s to $+60$~s around the alert time that would otherwise be overwritten due to lack of disk space and network bandwidth.

The DOM waveforms are digitized at 40~MHz, providing 25~ns time resolution for detected hits. Once the waveforms are requested and cached, they are available for offline analysis within 24 to 48~hours of the alert. The buffered requests thus overcome the limited 2~ms time resolution of the online counting-based MeV transient search. While this system was developed to enhance the analysis of self-triggered MeV neutrino transients, a data buffer request can also be externally triggered after the occurrence of an astrophysically important transient such as a gravitational wave alert \cite{IceCube:2023elowen} or a gamma-ray burst \cite{IceCube:2023grb} to enable prompt analysis.

\section{Searches for Galactic Transients with the MeV Neutrino Detection System}\label{sec:galactic}

At MeV, IceCube is very sensitive to short neutrino bursts originating in the immediate vicinity of the Milky Way. The largest signal would likely be produced by a core-collapse supernova (CCSN), which emits neutrinos with  luminosity $\mathcal{L}_\nu(t)$ exceeding $10^{51}$~erg~s$^{-1}$ (or more than $10^{56}$ neutrinos s$^{-1}$). During the core collapse, the neutrinos are emitted with a quasi-thermal spectrum parameterized in terms of average neutrino energy $\expval{E_\nu(t)}\approx10$ to $20~\mathrm{MeV}$ and a shape parameter $\alpha(t)$ \cite{Horiuchi:2018ofe}, where
\begin{align}\label{eq:ccsn_spectrum}
    f(E_\nu,t) & \propto \qty(\frac{E_\nu}{\expval{E_\nu}})^\alpha \exp{-\frac{(\alpha+1)E_\nu}{\expval{E_\nu}}},
    &
    \frac{\expval{E_\nu^k}}{\expval{E_\nu^{k-1}}} &\equiv \frac{2+\alpha}{1+\alpha}\expval{E_\nu},
    &
    k&=2,3,\ldots
\end{align}
Eq.~\eqref{eq:ccsn_spectrum} includes an implicit time dependence because $\expval{E_\nu}$ and $\alpha$ evolve significantly during the seconds after core bounce. Neutrino production also varies considerably depending on the mass of the progenitor and the physics of the explosion. During the collapse, complex and poorly-understood neutrino flavor transformations such as collective oscillations may occur in the core. Descriptions of the model dependence of CCSN neutrino emission can be found in \cite{Horiuchi:2018ofe, Burrows:2020qrp, SNEWS:2021ewj} and references therein.

In IceCube, the excess count rate per DOM produced by $\bar{\nu}_e$ IBD events from a supernova at a distance $d$ from Earth is given by
\begin{equation}\label{eq:ccsn_luminosity}
	r(t) \propto 
 \frac{n_\mathrm{target} \,\mathcal{L}_\mathrm{\nu}(t)}{4\pi d^2\expval{E_\nu(t)}} \int_0^{\infty} dE_{e^+} \int_0^{\infty} dE_\mathrm{\nu} \times \frac{d\sigma}{dE_\mathrm{e^+}}(E_{e^+},E_{\nu}) \,V_{e^+}^\mathrm{eff} \, f(E_{\nu},t)~~\qty[\mathrm{count\ s}^{-1}],
\end{equation}
where $n_\mathrm{target}$ is the number of proton targets in the ice and $f(E_{\nu},t)$ is the normalized $\bar{\nu}_e$ spectrum from eq.~\eqref{eq:ccsn_spectrum}. $V_{e^+}^\mathrm{eff}$ is the DOM effective volume for detecting an IBD positron, produced with cross section $d\sigma/dE_{e^+}$. The effective volume $V_{e^+}^\mathrm{eff}/E_{e^+}$ depends strongly on the optical scattering and absorption properties of the ice and the efficiency of the DOMs, ranging from 10~m$^3$~MeV$^{-1}$ in the main IceCube dust layer to 60~m$^3$~MeV$^{-1}$ for DeepCore DOMs in the clear ice near the bottom of the detector. 

Accounting for these effects, IceCube's effective mass for CCSN neutrinos emitted at a distance $d=10$~kpc (slightly further than the distance to the Galactic Center) is $\sim600$~kt per DOM. A CCSN at this location would produce $10^5$ to $10^6$ detected $\bar{\nu}_e$ events in IceCube. Given an observed excess count rate and an estimate of the spectral shape of the neutrinos, and assuming the distance $d$ and shape of the $\bar{\nu}_e$ spectrum are reasonably well known, we can estimate the $\bar{\nu}_e$ luminosity as
\begin{equation}\label{eq:ccsn_lum}
  \mathcal{L}_{\bar{\nu}_e}(t) \propto
  10^{47}\cdot\qty(\sum_{i=1}^{N_\mathrm{DOM}}r_i(t))\cdot
  \qty(\frac{d}{10~\mathrm{kpc}})^2\cdot
  \qty(\frac{\expval{E_\nu(t)}}{15~\mathrm{MeV}})^{-2}\cdot
  \qty(\frac{\qty(1+\alpha(t))^2}{\qty(2+\alpha(t))\qty(3+\alpha(t))})~~\qty[\mathrm{erg~s}^{-1}].
\end{equation}
%
%
%


Due to the aforementioned uncertainties in CCSN $\nu$ luminosity, progenitor distances, and fundamental neutrino interactions, we consider the sensitivity of IceCube to a wide variety of physical scenarios.
The fast open-source ASTERIA Monte Carlo code \cite{ASTERIA} is used to estimate the total count rate in all DOMs, weighted by DOM efficiency $\epsilon_i$, and the simulated counts are then injected atop background-only data measured online with the IceCube pulse-counting system. An example neutrino light curve for a simulated CCSN located at the Galactic Center is shown in \cite{IceCube:2023sn}. The simulation chain can be used to characterize the sensitivity of IceCube to \textsl{any} MeV neutrino transient such as novae or gravitational wave counterparts.

While IceCube requires external measurements of progenitor distance and the neutrino spectrum to fully constrain CCSN luminosity as in eq.~\eqref{eq:ccsn_lum}, the detector will provide a real-time high-statistics measurement of a CCSN in the Milky Way and the Magellanic Clouds. The excellent time resolution of the measurement (2~ms online and 25~ns offline) allows for detailed studies of the time structure of the neutrino emission during all phases of the burst. Further simulations of CCSNe in IceCube, estimates of the detector's sensitivity to a wide variety of core-collapse models as a function of distance, and the interaction of the detector with the SuperNova Early Warning System, are discussed in detail in \cite{IceCube:2023sn}.



\section{Constraining Extragalactic Transients with the MeV Detection System}\label{sec:extragalactic}

The DOM backgrounds and the small $\nu$ interaction cross section at MeV limit the discovery potential of the current IceCube pulse-counting system to transients in the Milky Way and Magellanic Clouds. As a result, we are constrained to place upper limits on extragalactic neutrino emission for most transients. However, an advantage of using IceCube data is the excellent stability and high duty cycle of the detector, whose live time exceeds 99.7\% per year. This means IceCube is highly likely to be in stable data acquisition during most astrophysical transients of interest.

To construct upper limits, we look for excess counts in the data during a time window determined by the transient. We proceed using two approaches:
\begin{enumerate}
    \setlength\itemsep{-0.5em}
    \item Assume a quasi-thermal MeV emission spectrum for the neutrino emission and scale the well-characterized sensitivity to CCSNe to compute a 90\% U.L. on $\mathcal{L}_{\bar{\nu}_e}(t)$ using eq.~\eqref{eq:ccsn_lum}.
    \item Compute delta-function mono-energetic limits on $\mathcal{L}_{\bar{\nu}_e}(t)$ and construct a model-independent upper limit on luminosity as a function of $E_\nu$.
\end{enumerate}
The first approach has been used in previous IceCube analyses, such as searches for MeV emission from Fast Radio Bursts (FRBs) \cite{IceCube:2019acm} and GRB~221009A \cite{IceCube:2023rhf}. 

The FRB analysis, for example, used the procedure described in Sec.~\ref{sec:online_system}, but instead of searching in 0.5~s time windows, signal bins of widths ranging from 10~ms to 1280~ms we used to bracket the times of known FRBs. The background distributions $\expval{r}_i$ and $\sigma_i$ were computed using off-source background-only 10-second windows sampled during an 8-hour period around the detection of the FRB, and these were used to construct an estimated excess count rate $\Delta\mu$ using eq.~\eqref{eq:delta_mu} and the distribution of the muon-corrected test statistic $\xi_\mathrm{corr}$ under the null hypothesis. From $\xi_\mathrm{corr}$, a 90\% U.L. on the neutrino flux could be computed using count rates measured in the FRB search windows.

The analysis of MeV neutrinos from GRB~221009A proceeded in a similar fashion, using six search windows motivated by models of neutrino-dominated accretion flows, a boosted shock breakout precursor, and a GRB fireball model (see details in \cite{IceCube:2023grb, IceCube:2023rhf}). The ASTERIA fast Monte Carlo was used to compute the sensitivity of IceCube to different MeV emission models for the GRB, all assuming a quasi-thermal spectrum similar to eq.~\eqref{eq:ccsn_spectrum}. The pulse count rates observed in all search windows were consistent with backgrounds measured in off-source windows of equivalent size. As a result, IceCube reported 90\% upper limits on the MeV $\bar{\nu}_e$ flux for each of the six search windows \cite{IceCube:2023rhf}.

\begin{figure}[ht]
    \includegraphics[width=\textwidth]{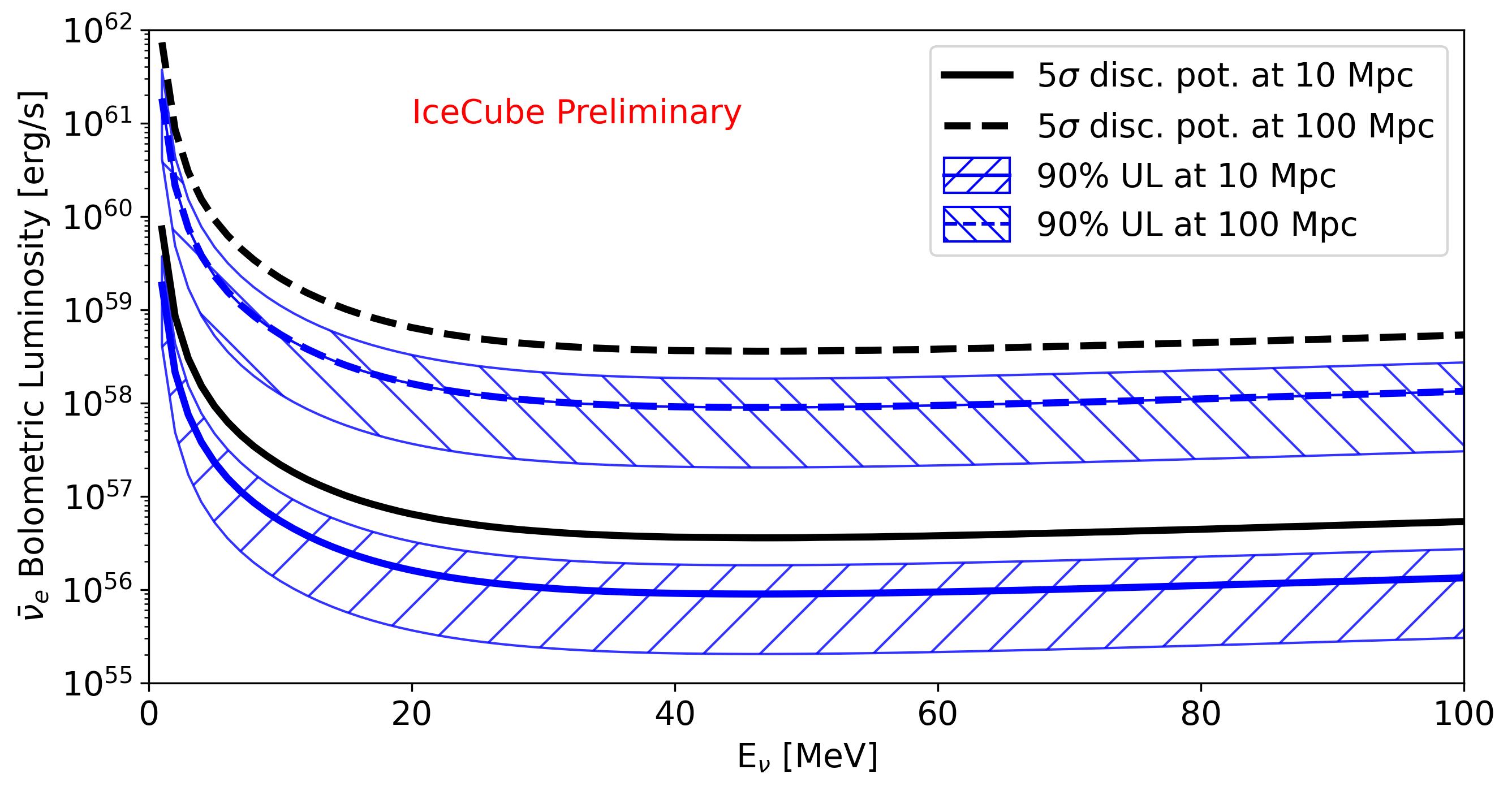}
    \caption{The quasi-differential sensitivity of IceCube to MeV $\bar{\nu}_e$ emission from a transient at 10~Mpc (solid blue curve) and 100~Mpc (dashed blue curve) using a 1~s search window. At both distances, we plot the median 90\% upper limits expected for the bolometric luminosity (central blue curves), while the hatched regions indicate the central 68\% confidence interval on the upper limits. We also indicate the $5\sigma$ discovery potential for IceCube for a source at 10~Mpc (solid black line) and 100~Mpc (dashed black line).}
    \label{fig:exgal_sensitivity}
\end{figure}

A limitation of the approach used in the FRB and GRB searches is the assumption of a thermal or quasi-thermal spectrum. For cases where there is no generally accepted MeV neutrino emission model, we adopt the non-parametric approach described in \cite{Super-Kamiokande:2002zkw, Turcan:2003yf}. Here, we compute the flux limit corresponding to a delta function spectrum in energy, using ASTERIA to simulate a mono-energetic flux from a source at a given distance for a specific energy. The process is then repeated in steps of $1$~MeV for $E_\nu$ between $1$ and $100$~MeV.


Fig.~\ref{fig:exgal_sensitivity} shows the expected sensitivity of IceCube to extragalactic MeV neutrino emission from a source at 10~Mpc and 100~Mpc from Earth. A search window of 1~s is assumed, within which we compute the maximum test statistic for the count rate in two bins of width 0.5~s. The distribution of the test statistic is constructed using background-only simulations. An independent sample of equivalent background-only simulations is used to estimate a 90\% upper limit on the count rate (and $\mathcal{L}_{\bar{\nu}_e}$ given the fixed distance) at a fixed value $E_\nu$. Using the test sample, we estimate the median 90\% upper limit and the central 68\% interval around the median for fixed energy $E_\nu$. The process is repeated for all energies between 1 and 100~MeV for sources at 10 and 100~Mpc to construct the distributions as a function of $E_\nu$.


Fig.~\ref{fig:exgal_sensitivity} also shows the potential of IceCube to identify $5\sigma$ evidence for MeV neutrino emission from an extragalactic transient. The ASTERIA code was used to estimate the mono-energetic $\bar{\nu}_e$ luminosity that produces a count rate in the $5\sigma$ upper tail of the test statistic constructed for this search in at least 50\% of simulated search windows. The calculation was then repeated for all energies between 1 and 100~MeV. The threshold luminosity is quite large, meaning that IceCube would only detect evidence from a super-luminous transient, but it comes with the advantages of being independent of assumptions about the neutrino spectrum and can be applied to any transient observed while IceCube is in data acquisition.

\begin{figure}[ht]
    \includegraphics[width=\textwidth]{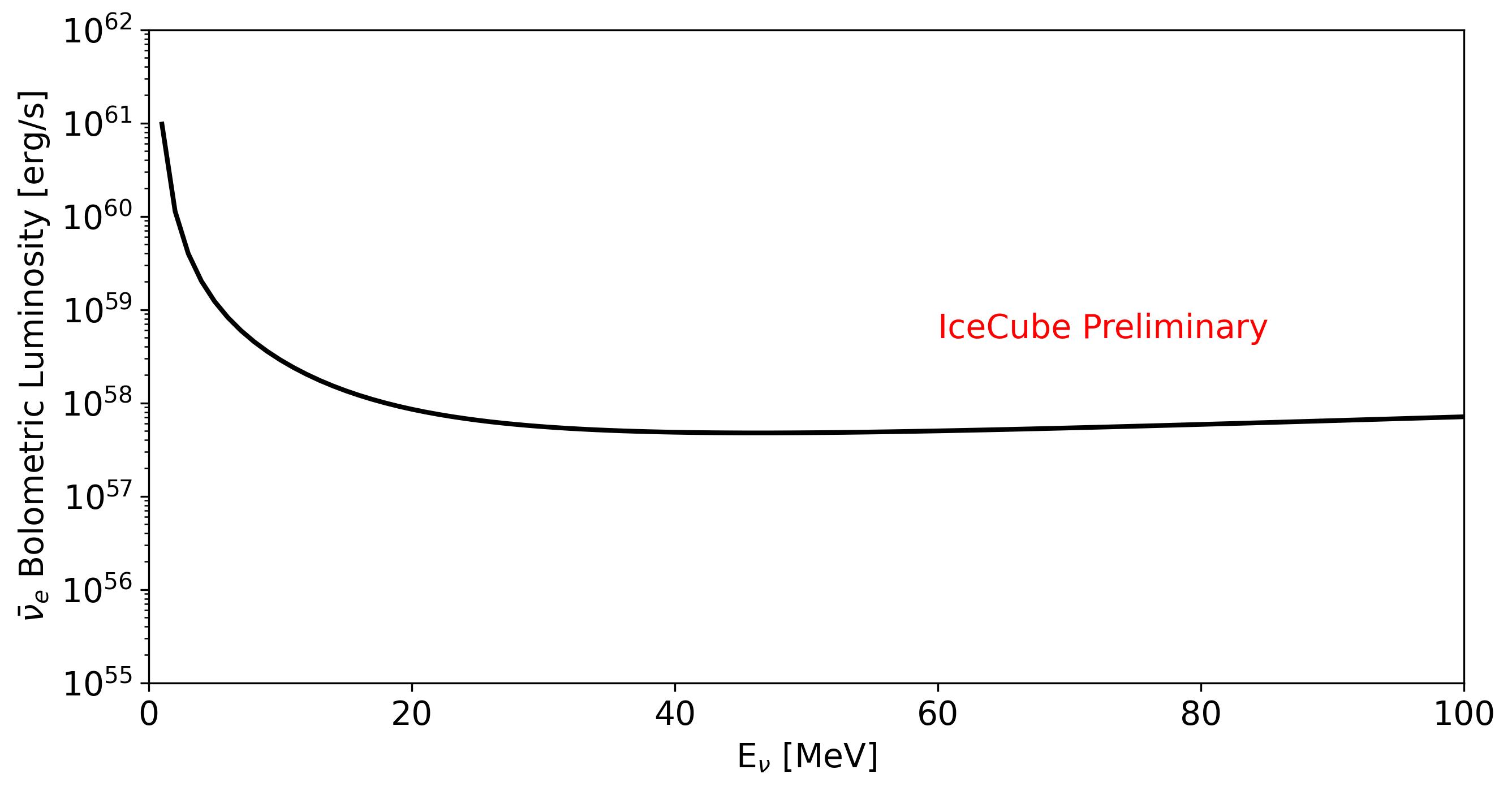}
    \caption{90\% upper limit on the $\bar{\nu}_e$ luminosity of SN 2023ixf, a CCSN located 6.4~Mpc from Earth that was discovered on May 19, 2023. }
    \label{fig:ul_sn2023ixf}
\end{figure}

An example limit is shown in Fig.~\ref{fig:ul_sn2023ixf} using SN~2023ixf, a Type II CCSN discovered in M101 on May 19, 2023 \cite{Itagaki:2023}. IceCube was in data acquisition prior to the time of the discovery. Since no neutrino burst was observed and we had no knowledge of the time of the core collapse, we placed a conservative upper limit on the MeV $\bar{\nu}_e$ luminosity by using the largest upward fluctuation in scalar counts in 0.5~s bins recorded during the five days leading up to the discovery.



\section{Conclusions}\label{sec:conclusions}

IceCube, which is designed to reconstruct neutrino events from several GeV to PeV, also provides a large-volume and high-uptime monitor for MeV transients lasting a few seconds. We have presented sensitivities for model-dependent searches for supernova neutrinos and other transients with quasi-thermal energy spectra, as well as model-independent sensitivities that do not depend on assumptions about the spectrum. The backgrounds in the current IceCube + DeepCore DOMs limit the realistic discovery potential to Galactic transients at MeV, but IceCube's live time does allow for calculations of upper limits on MeV $\nu$ emission from most fast transients in the nearby universe. Future detectors deployed at IceCube will include multi-PMT DOMs which will enable substantial reductions in the DOM backgrounds for short-timescale analyses \cite{LozanoMariscal:2021the}, similar to the capabilities of detectors such as KM3Net \cite{Molla:2019scm}. The improvements of these upgraded sensors on the sensitivity and discovery potential for detecting MeV transients will be the subject of future work.

\bibliographystyle{ICRC}
\bibliography{references}

\clearpage
%
%
%

\input{authorlist_icecube.tex}

\end{document}

%% file: authorlist_icecube.tex
\section*{Full Author List: IceCube Collaboration}

\scriptsize
\noindent
R. Abbasi$^{17}$,
M. Ackermann$^{63}$,
J. Adams$^{18}$,
S. K. Agarwalla$^{40,\: 64}$,
J. A. Aguilar$^{12}$,
M. Ahlers$^{22}$,
J.M. Alameddine$^{23}$,
N. M. Amin$^{44}$,
K. Andeen$^{42}$,
G. Anton$^{26}$
% ...
\\
\\
% ...
$^{12}$ Universit{\'e} Libre de Bruxelles, Science Faculty CP230, B-1050 Brussels, Belgium \\
% ...
$^{17}$ Department of Physics, Loyola University Chicago, Chicago, IL 60660, USA \\
$^{18}$ Dept. of Physics and Astronomy, University of Canterbury, Private Bag 4800, Christchurch, New Zealand \\
% ...
$^{22}$ Niels Bohr Institute, University of Copenhagen, DK-2100 Copenhagen, Denmark \\
$^{23}$ Dept. of Physics, TU Dortmund University, D-44221 Dortmund, Germany \\
% ...
$^{26}$ Erlangen Centre for Astroparticle Physics, Friedrich-Alexander-Universit{\"a}t Erlangen-N{\"u}rnberg, D-91058 Erlangen, Germany \\
% ...
$^{40}$ Dept. of Physics and Wisconsin IceCube Particle Astrophysics Center, University of Wisconsin{\textendash}Madison, Madison, WI 53706, USA \\
% ...
$^{42}$ Department of Physics, Marquette University, Milwaukee, WI, 53201, USA \\
% ...
$^{44}$ Bartol Research Institute and Dept. of Physics and Astronomy, University of Delaware, Newark, DE 19716, USA \\
% ...
$^{63}$ Deutsches Elektronen-Synchrotron DESY, Platanenallee 6, 15738 Zeuthen, Germany  \\
$^{64}$ Institute of Physics, Sachivalaya Marg, Sainik School Post, Bhubaneswar 751005, India \\
% ...

\clearpage
\subsection*{Acknowledgments}

\noindent
The authors gratefully acknowledge the support from the following agencies and institutions:
USA {\textendash} U.S. National Science Foundation-Office of Polar Programs,
U.S. National Science Foundation-Physics Division,
U.S. National Science Foundation-EPSCoR,
U.S. National Science Foundation-Office of Advanced Cyberinfrastructure,
Wisconsin Alumni Research Foundation,
Center for High Throughput Computing (CHTC) at the University of Wisconsin{\textendash}Madison,
Open Science Grid (OSG),
Partnership to Advance Throughput Computing (PATh),
Advanced Cyberinfrastructure Coordination Ecosystem: Services {\&} Support (ACCESS),
Frontera computing project at the Texas Advanced Computing Center,
U.S. Department of Energy-National Energy Research Scientific Computing Center,
Particle astrophysics research computing center at the University of Maryland,
Institute for Cyber-Enabled Research at Michigan State University,
Astroparticle physics computational facility at Marquette University,
NVIDIA Corporation,
and Google Cloud Platform;
Belgium {\textendash} Funds for Scientific Research (FRS-FNRS and FWO),
FWO Odysseus and Big Science programmes,
and Belgian Federal Science Policy Office (Belspo);
Germany {\textendash} Bundesministerium f{\"u}r Bildung und Forschung (BMBF),
Deutsche Forschungsgemeinschaft (DFG),
Helmholtz Alliance for Astroparticle Physics (HAP),
Initiative and Networking Fund of the Helmholtz Association,
Deutsches Elektronen Synchrotron (DESY),
and High Performance Computing cluster of the RWTH Aachen;
Sweden {\textendash} Swedish Research Council,
Swedish Polar Research Secretariat,
Swedish National Infrastructure for Computing (SNIC),
and Knut and Alice Wallenberg Foundation;
European Union {\textendash} EGI Advanced Computing for research;
Australia {\textendash} Australian Research Council;
Canada {\textendash} Natural Sciences and Engineering Research Council of Canada,
Calcul Qu{\'e}bec, Compute Ontario, Canada Foundation for Innovation, WestGrid, and Digital Research Alliance of Canada;
Denmark {\textendash} Villum Fonden, Carlsberg Foundation, and European Commission;
New Zealand {\textendash} Marsden Fund;
Japan {\textendash} Japan Society for Promotion of Science (JSPS)
and Institute for Global Prominent Research (IGPR) of Chiba University;
Korea {\textendash} National Research Foundation of Korea (NRF);
Switzerland {\textendash} Swiss National Science Foundation (SNSF).